\documentclass[%
 aip,
 amsmath,amssymb,
 reprint,%
]{revtex4-1}
\usepackage{dcolumn}
\usepackage{bm}
\usepackage[utf8]{inputenc}
\usepackage[T1]{fontenc}
\usepackage{mathptmx} 
\usepackage[normalem]{ulem}
\usepackage{mathtools}
\usepackage{graphicx}
\usepackage{epstopdf}
\usepackage{lmodern}
\usepackage{xfrac}
\usepackage[breaklinks]{hyperref}
\usepackage{mathtools}
\usepackage{xcolor}
\newcommand{\ket}[1]{|{#1}\rangle}
\newcommand{\bra}[1]{\langle{#1}|}
\newcommand*\diff{\mathop{}\!\mathrm{d}}

\begin{document}

\title{Asymmetric delay attack on an entanglement-based bidirectional clock synchronization protocol}

\author{Jianwei~Lee}
\affiliation{Centre for Quantum Technologies, National University of
  Singapore, 3 Science Drive 2, Singapore 117543, Singapore}

\author{Lijiong~Shen}
\affiliation{Centre for Quantum Technologies, National University of
  Singapore, 3 Science Drive 2, Singapore 117543, Singapore}
\affiliation{Department of Physics, National University of Singapore, 2
  Science Drive 3, Singapore 117551, Singapore}

\author{Alessandro~Cer\`{e}}
\affiliation{Centre for Quantum Technologies, National University of
  Singapore, 3 Science Drive 2, Singapore 117543, Singapore}

\author{James~Troupe}
\affiliation{Applied Research Laboratories, The University of Texas at Austin,
  Austin, Texas, USA}

\author{Antia~Lamas-Linares}
\affiliation{SpeQtral, 73 Science Park Drive, Singapore 118254, Singapore}
\affiliation{Centre for Quantum Technologies, National University of
  Singapore, 3 Science Drive 2, Singapore 117543, Singapore}

\author{Christian~Kurtsiefer}
\affiliation{Centre for Quantum Technologies, National University of
  Singapore, 3 Science Drive 2, Singapore 117543, Singapore}
\affiliation{Department of Physics, National University of Singapore, 2
  Science Drive 3, Singapore 117551, Singapore}

\email[]{christian.kurtsiefer@gmail.com}
\date{\today}
\begin{abstract} 
We demonstrate an attack on a clock synchronization protocol that attempts to detect tampering of the synchronization channel using polarization-entangled photon pairs.
The protocol relies on a symmetrical channel, where propagation delays do not depend on propagation direction, for correctly deducing the offset between clocks --
a condition that could be manipulated with optical circulators, which rely on
static magnetic fields to break the reciprocity of propagating electromagnetic fields. 
Despite the polarization transformation induced within a set of circulators, our attack creates an error in time synchronization while successfully evading detection. 
\end{abstract}
\maketitle

\section{Introduction}
Clock synchronization protocols that bidirectionally exchange signals, 
e.g., the Network Time Protocol (NTP) or the two-way satellite time transfer (TWSTFT),  
are widely used to estimate the absolute time offset between remote clocks 
without first characterizing network propagation times~\cite{Mills:1991,PTP,piester2008time,jiang2017bipm}. 
By assuming that propagation delays are symmetric in the two directions of travel in a synchronization channel,
parties estimate one-way propagation times as half of the round-trip time.
Although convenient, this assumption exposes the protocol to attacks that
introduce unknown asymmetric channel delays
which cannot be detected by better encryption or authentication~\cite{narula:17}.
Existing countermeasures~\cite{mizrahi2012game,ullmann2009delay,tsang2006security}
e.g. based on monitoring round-trip times 
have been evaded by sophisticated intercept, spoofing and delay techniques~\cite{rabadi2017taming}. 

Recently, protocol implementations using entangled photons
suggest measuring non-local properties to ensure that synchronization networks have not been tampered with 
-- a technique associated with entanglement-based quantum key distribution~\cite{lee2019symmetrical,hou2018fiber,lamas2018secure}.
Tight time correlations between entangled photons prepared by spontaneous
parametric down conversion (SPDC) allow synchronizing independent atomic clocks at photon rates of order 100 pairs/s~\cite{lee2019symmetrical} and with potential accuracies $<1\,$ps~\cite{hou2018fiber}.
Monogamy of entanglement ensures that a counterfeit photon entangled with the legitimate signal cannot be generated, allowing signal authentication~\cite{yang2006simple}. 
The no-cloning theorem prevents intercept, copy and resend of an identical quantum state with an arbitrary delay~\cite{wootters1982single}. 

Despite these security enhancements,
the vulnerability to an asymmetric delay attack remains
since photons traveling in opposite directions can be passively rerouted with a circulator (Figure~\ref{fig:setup}) by using the Faraday effect to break the reciprocity of the channel. 
A recent proposal suggests that even polarization-insensitive circulators, 
which rotate input polarizations back to the same state, 
impose a measurable change in the phase of the joint state~\cite{Troupe:2018}. 
The proposal was based on the fact that the phase change after a cyclic
quantum evolution is measurable under certain conditions~\cite{berry:87}.
Previous experiments with entangled
photons~\cite{kwiat:91,strekalov:97,brendel:95,jha:08} seemed to support this
proposed protection.

In this work, we examine the circulator-based asymmetric delay attack~\cite{Troupe:2018}. 
We experimentally show that the attack \textit{cannot} be detected by the proposed mechanism and demonstrate an induced error in synchronization of over $25$\,ns between two rubidium clocks. 

\section{Attacking an Entanglement-Based Clock Synchronization Protocol}
We briefly review the clock synchronization protocol considered~\cite{Troupe:2018}.
\begin{figure}
\begin{center}
\includegraphics[width=\linewidth]{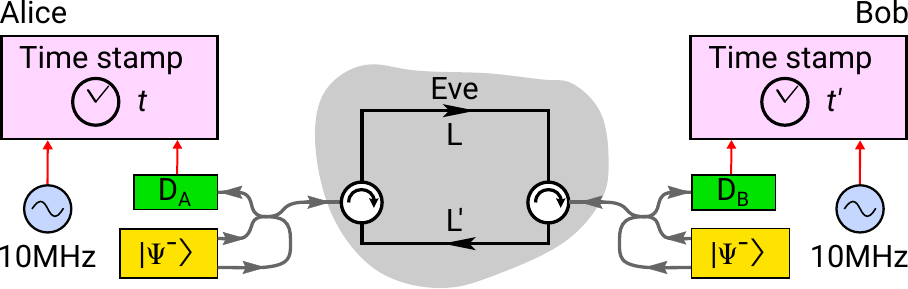}
   \caption{\label{fig:setup}
    Clock synchronization scheme. 
    Alice and Bob each have a source of polarization-entangled photon pairs $\ket{\Psi^-}$, 
    and avalanche photodetectors at D$_{\text{A,B}}$. 
    One photon of the pair is detected locally, 
    while the other
    photon is sent through a fiber to be detected on the remote side. 
    Arrival times for all detected photons are recorded at each side with respect to local clocks, 
    each locked to a rubidium frequency reference.
    Grey region: asymmetric delay attack.
    An adversary (Eve) 
    uses a pair of circulators to introduce a direction-dependent propagation
    delay:
    photons originating at Bob’s site
    will always take the bottom path, while photons originating at
    Alice’s side will take the top path. 
     }
\end{center}
\end{figure}
The protocol involves two parties, Alice and Bob, connected by a single mode optical channel. 
Each party has a source of polarization-entangled photons pairs generated by SPDC. 
One photon of the pair is detected locally, while the other is sent and detected on the remote side~(Figure~\ref{fig:setup}). 
Every photodetection event is time-tagged with respect to a local clock which assigns time stamps $t$ and $t'$.  

Photon pairs emerging from SPDC are tightly time-correlated.
Thus, for an offset $\delta$ between the clocks,  
a propagation time $\Delta t_{AB}$ from Alice to Bob, 
and $\Delta t_{BA}$ in the other direction, 
the second-order correlation function $G^{(2)}(\tau = t'-t)$ of the time difference
has two peaks at
\begin{equation}
  \tau_{AB} = \delta + \Delta t_{AB} \quad\mathrm{and}\quad
  \tau_{BA} = \delta - \Delta t_{BA}\,
\end{equation}
due to pairs created by Alice and Bob~\cite{glauber1963quantum}.
A round-trip time $\Delta T$ for photons can be calculated using the inter-peak separation,
\begin{equation}\label{eq:round_trip}
\Delta T = \Delta t_{AB} + \Delta t_{BA} = \tau_{AB}-\tau_{BA},
\end{equation}
while the offset
\begin{equation}\label{eqn:inaccurate_offset}
  \delta = \frac{1}{2}\,[(\tau_{AB} + \tau_{BA}) - (\Delta t_{AB} - \Delta t_{BA})]
\end{equation} 
is given by the midpoint of the peaks and a propagation delay asymmetry, respectively.
Assuming a symmetrical propagation delay, 
$\Delta t_{AB}=\Delta t_{BA}$, 
the clock offset 
\begin{equation}\label{eq:offset}
\delta = \frac{1}{2}\left (\tau_{AB}+\tau_{BA}\right )
\end{equation}
is obtained directly from the midpoint.

Eve may now may exploit this assumption by separating the two propagation directions with a pair of circulators (Figure~\ref{fig:setup} gray region), introducing a direction dependent delay 
$\Delta t_{AB} - \Delta t_{BA} = (L-L')/v$,
where $L$ is the additional propagation length from Alice to Bob, and $L'$ in
the other direction, and $v$ is the speed of light in the fiber.
If Alice and Bob continue to rely on the midpoint between the peaks to estimate $\delta$, they will obtain instead $\delta + (L-L')/2v$.

In an attempt to detect the circulators, 
Ref.~\onlinecite{Troupe:2018} suggests that Alice and Bob 
monitor polarization correlations using avalanche photodiode preceded by a polarization measurement in the appropriate bases (D$_{\text{A,B}}$). 
The detection scheme is based on the fact that circulators use Faraday Rotation to separate photons propagating in opposite directions - 
Faraday Rotation is a time-reversal symmetry breaking mechanism that rotates polarization, potentially changing the input state. 

For each individual polarization state to be preserved, the circulators must
rotate the state by an integer multiple of 180${^{\text{o}}}$
so that for a Bell state $\ket{\Psi^-} = \frac{1}{\sqrt{2}}\left(\ket{HV} - \ket{VH}\right)$ distributed by Alice,
the rotation of Bob's state ($\ket{\psi}_B\rightarrow\pm\ket{\psi}_B$) does not result in any measurable change
\begin{equation}\label{eqn:direct_result}
\ket{\Psi^-} \rightarrow \pm\frac{1}{\sqrt{2}}\left(\ket{HV} - \ket{VH}\right) = \pm\ket{\Psi^-}.
\end{equation}
However, as the evolution of Bob's state follows a closed trajectory on the
Poincaré sphere, Ref.~\onlinecite{Troupe:2018} predicted that a geometric
phase -- the phase determined by the geometry of the trajectory on the
sphere~\cite{berry:87} -- is imposed on the Bell state, and can be detected in
a non-local measurement.  
We show in the supplementary material that 
when other phase contributions are taken into account, 
the net effect of the circulators nonetheless produce no measurable change to the Bell state (Eq.~\ref{eqn:direct_result}). 
We use this result and experimentally demonstrate a successful asymmetric delay attack using the circulators in subsequent sections. 

\section{Experiment}\label{sec:experiment}
We first implement the clock synchronization protocol.
For two independent rubidium clocks, the following setup was previously characterized to achieve a synchronization precision of 51\,ps in 100\,s, comparable to the relative intrinsic frequency instability of each clock~\cite{lee2019symmetrical}.

Two identical SPDC sources generate polarization-entangled photon pairs (Figure~\ref{fig:setup}).
The output of a laser diode (power $\approx$10\,mW, central wavelength 405\,nm) is coupled into a single mode optical fiber (SMF) for spatial mode filtering and focused to a beam waist of 80\,$\mu$m into a 2\,mm thick $\beta$-Barium~Borate crystal cut for non-collinear type-II phase matching~\cite{Kwiat:1995ub}.
Down-converted photons at 810\,nm are coupled into two single
mode fibers with an overall detected pair rate of about~$200$\,s$^{-1}$.
Fiber beam splitters separate the photon pairs so that one photon is detected locally with an avalanche photodetector (D$_{\text{A,B}}$), while the other photon is transmitted to the remote party. 

Time-stamping units assign detection times
$t$ and $t'$ to the events detected at Alice and Bob, respectively. 
We compute the histogram $G^{(2)}(\tau = t' - t)$ of the time differences and resolve two coincidence peaks (FWHM $\approx$ 500\,ps) with a resolution of 16\,ps, one from each source~\cite{ho2009clock}.
The offset and round-trip-times are determined from the mean and separation of the peaks, respectively.
For the purposes of this demonstration, we lock the clocks with unknown offset to a common rubidium frequency reference, thus avoiding frequency drifts that can detract from the main point of the experiment, i.e. demonstrating an induced error in offset estimation.

\subsection{Asymmetric Delay Attack}
To implement the asymmetric delay attack, 
we use two 3-port polarization-insensitive optical circulators of design-wavelength 810\,nm 
and two single mode fibers of lengths $L$ and $L'$.
\begin{figure}
\begin{center}
\includegraphics[width=\linewidth]{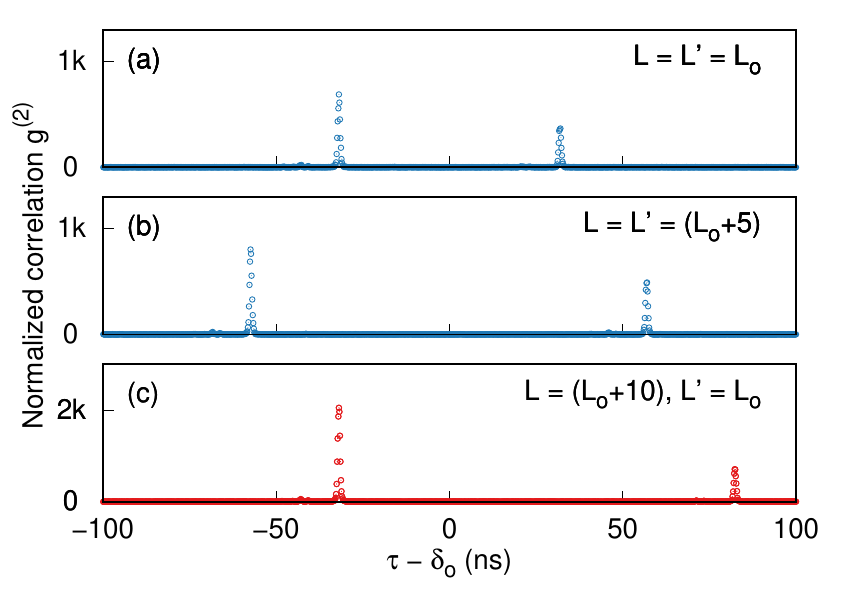} 
  \caption{\label{fig:g2}
  Time correlations of Alice and Bob's detection events normalized to background coincidences.
  The separation between peaks corresponds to the round-trip time~$\Delta T$, and the midpoint is the offset between the clocks~$\delta$.
  Symmetric delays with $L=L'$ show that the offset remains constant for both the (a) initial and (b) extended round-trip times.
  An asymmetric delay with (c) $L=L'+10$ results in an offset shift.
  $L_o/2$: minimum length of the fiber belonging to each circulator port.   
  ${\delta_{\text{o}}}$: the offset estimated in (a).
}
\end{center}
\end{figure}
\begin{figure}
\begin{center}
\includegraphics[width=\linewidth]{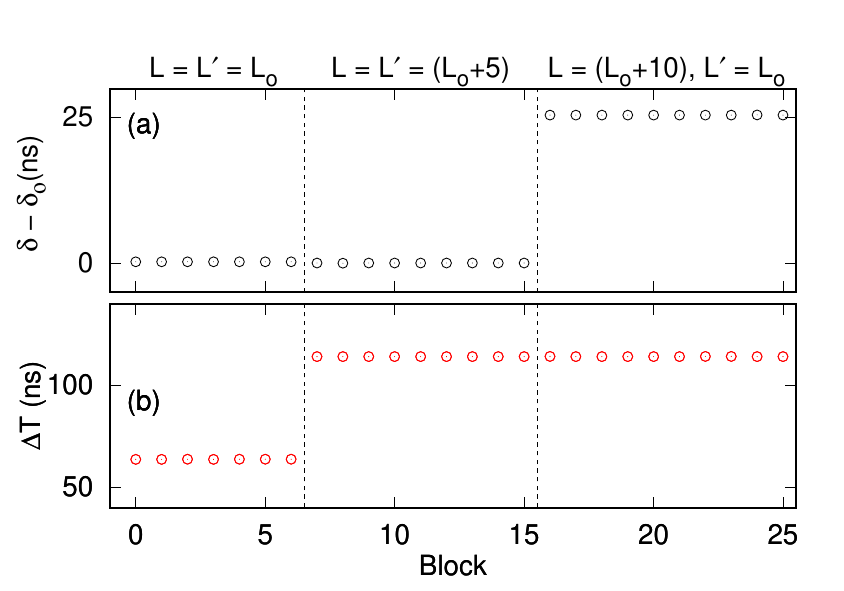}
  \caption{\label{fig:asymmetric_attack_results}
  (a) Measured offset $\delta$ between two clocks, both locked on the same frequency reference. 
  Each value of $\delta$ was evaluated from measuring photon pair timing correlations from a block of photodetection times recorded by Alice and Bob. Each block is 40\,s long.
  (b) The round-trip time $\Delta T$.
  Block 6 to 7: increasing the symmetric delay ($L=L'$) does not change $\delta$. 
  Block 15 to 16: introducing an asymmetric delay ($L\neq L'$) creates an offset error.
  $\delta_{\text{o}}$: offset measured in the first block.
}
\end{center}
\end{figure}

We first estimate the initial offset ${\delta_{\text{o}}}$ between the two clocks with a symmetric channel delay $L=L'=L_{\text{o}}$.
Figure~\ref{fig:g2}(a) shows $g^{(2)}(\tau)$, the second-order correlation function $G^{(2)}(\tau)$ normalized to background coincidences, acquired from the time stamps recorded for about 5\,min. In Figure~\ref{fig:asymmetric_attack_results} we plot the offset and round-trip times estimated every 40\,s.

To illustrate the difference in the cross-correlation measured between a symmetric and an asymmetric delay attack, 
we use two 5\,m fibers to impose an additional round-trip of 10\,m, but distribute them differently during each attack.
For the symmetric delay attack, we extend $L$ and $L'$ equally by 5\,m.
We observe in Figure~\ref{fig:g2}(b) that although the peak separation increases, the midpoint of the peaks used for estimating the offset remains unchanged.
For the asymmetric delay attack, both fibers are used to extend $L$ by 10\,m, while $L'$ remains unchanged. 
We observe in Figure~\ref{fig:g2}(c) that the peak separation remains the same
as in Figure~\ref{fig:g2}(b), but the midpoint of the peaks has shifted by
$25.24(2)$\,ns corresponding to half the additional round-trip time
incurred. This indicates a successful attack.

\subsection{Asymmetric Delay Attack Detection}
\begin{figure}
\begin{center}
\includegraphics[width=1\linewidth]{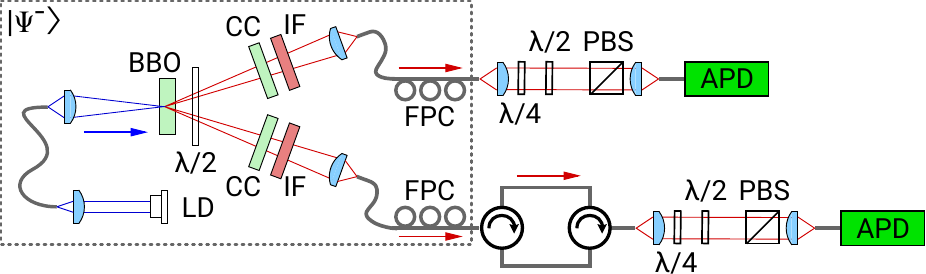}
  \caption{\label{fig:tomography_setup}
Setup for quantum state tomography on a polarization-entangled photon pair state,
with one photon passing through a pair of circulators. 
Dashed box: optical setup of our polarization-entangled photon source~\cite{Kwiat:1995ub}. 
LD: laser diode, BBO: $\beta$-Barium~Borate, CC: compensation crystals, FPC: fiber polarization controller, SMF: single mode fiber, $\lambda$/4: quarter-wave plate, $\lambda$/2: half-wave plate, PBS: polarizing beam splitter, APD: avalanche photodiode.
     }
\end{center}
\end{figure}
As a proof-of-principle demonstration of how the circulators influence the distributed entanglement, 
we measure polarization correlations of Alice's pair source before and after the circulators are inserted in one of its output modes with the setup shown in Figure~\ref{fig:tomography_setup}.
For each output mode, a quarter-wave plate (QWP), half-wave plate (HWP) and polarizing beamsplitter (PBS) projects the polarization mode into either $\ket{H}, \ket{V}, \ket{D}, \ket{A}, \ket{L} \text{ or } \ket{R}$.
Fiber polarization controllers (FPCs) correct for the polarization errors introduced by the fibers.
We note that since FPCs do not break time-reversal symmetry, they cannot invert the polarization transformation induced by the circulators.
We detect photon pairs with APDs for 36 wave plate settings 
and numerically search for the density matrix most likely to have returned the observed pair rates~\cite{altepeter2005photonic}.

\begin{figure}
    \centering
        \includegraphics[width=\columnwidth]{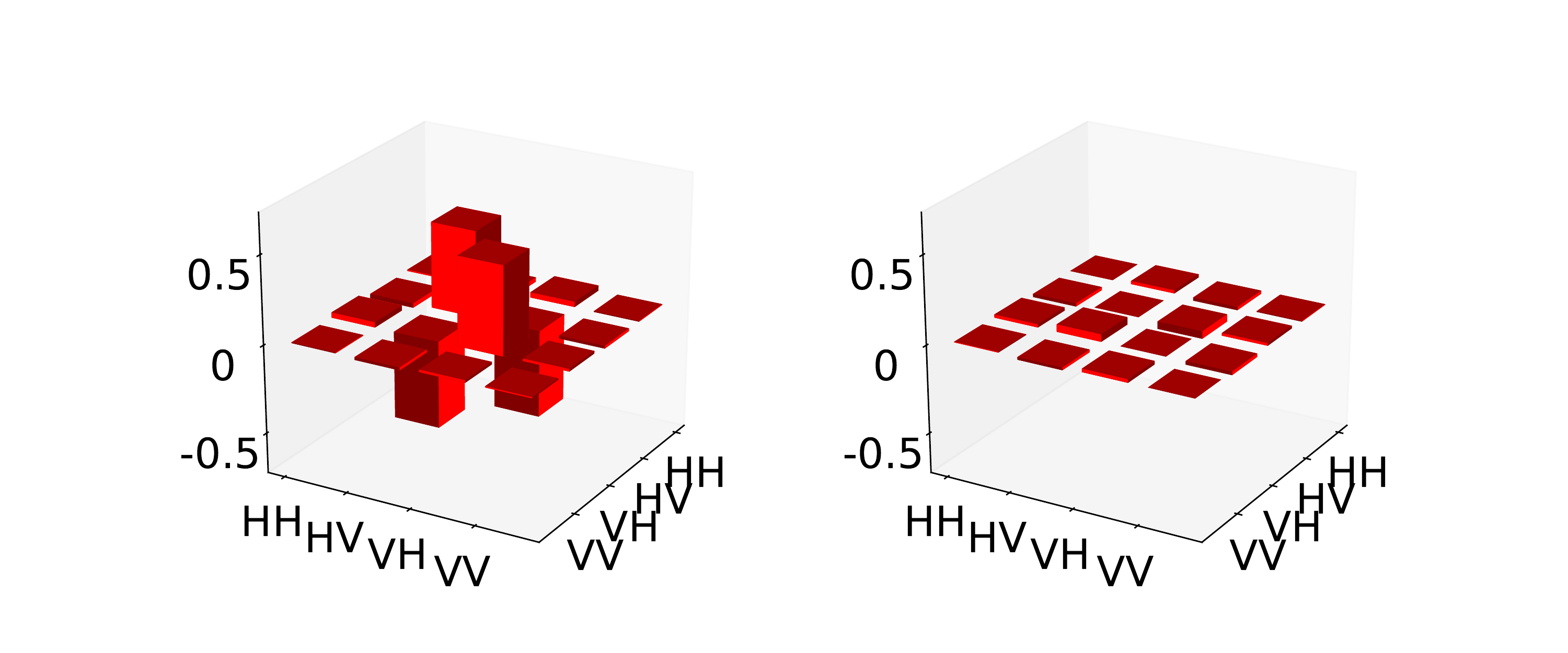}
        (a) Before insertion of circulators, fidelity with $\ket{\Psi^-}$: $98.2\%$.

        \includegraphics[width=\columnwidth]{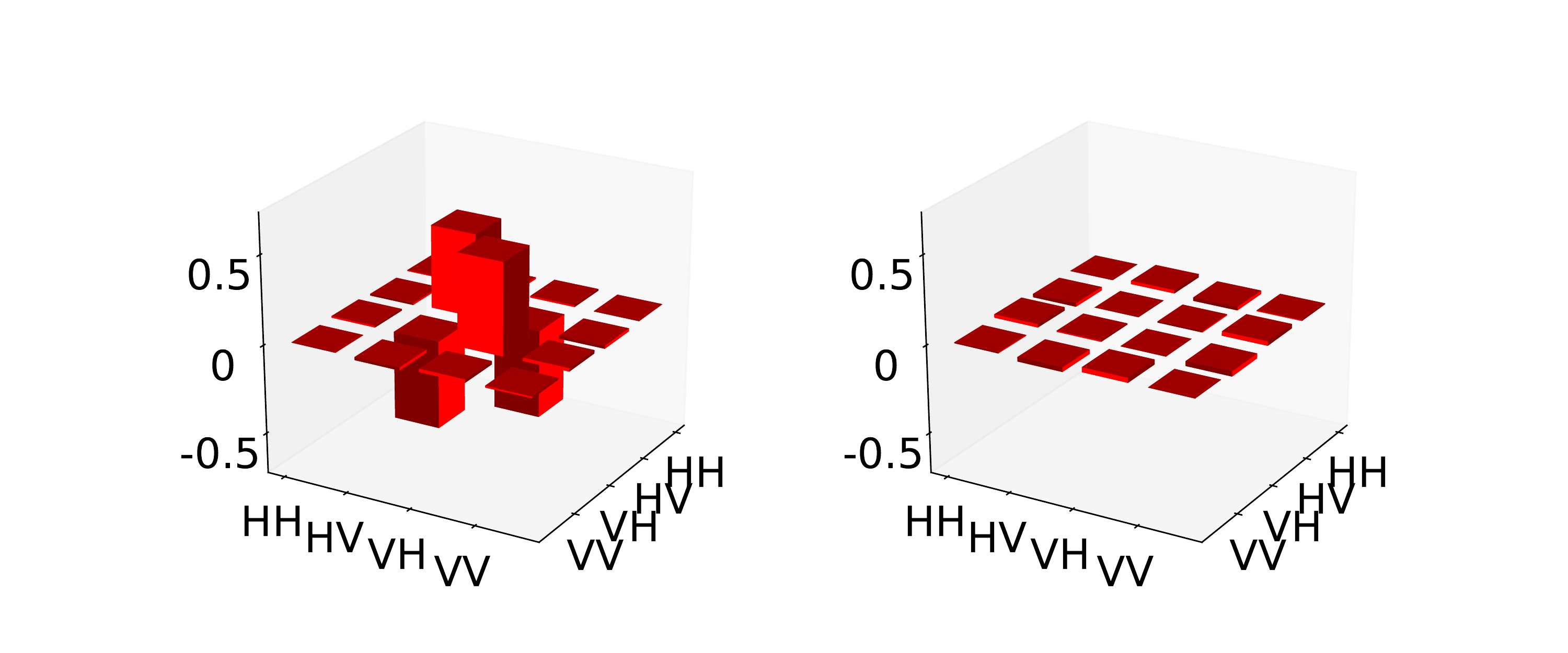}
        (b) After insertion of circulators, fidelity with $\ket{\Psi^-}$: $98.4\%$.
    \caption{\label{fig:rho}
    Real and imaginary part of the reconstructed density matrix for the target Bell state $\ket{\Psi^-}$ originating from Alice's source. 
    Bob receives one photon of the pair through the synchronization channel.
    The density matrices obtained (a) before and (b) after
    polarization-insensitive circulators are inserted
    (Figure~\ref{fig:tomography_setup}) do not deviate significantly from
    $\ket{\Psi^-}$.
    }
\end{figure}

Figure~\ref{fig:rho} shows the reconstructed density matrices of Alice's state before ($\rho_\text{o}$) and after ($\rho$) the introduction of the circulators into the path of Bob's photons.
\begin{figure}
\begin{center}
\includegraphics[width=0.92\linewidth]{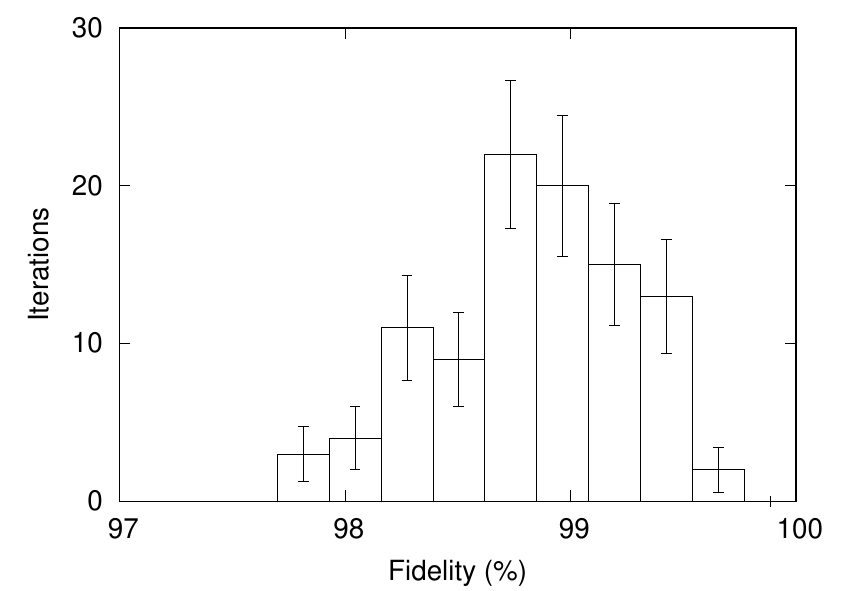}
  \caption{\label{fig:fidelity_distribution}
  Fidelity distribution comparing the Bell state originating from Alice's source before and after introducing the circulators. 
  The distribution is generated by numerically propagating errors due to counting statistics. 
  A high mean fidelity suggests that the state remains unchanged and cannot be used to detect the attack. 
  Error bars: Poissonian standard deviation.
     }
\end{center}
\end{figure}
We compare $\rho_\text{o}$ and $\rho$ by computing the fidelity  $F(\rho,\rho_\text{o})=\left(\textrm{Tr}\sqrt{\sqrt{\rho}\rho_\text{o}\sqrt{\rho}}\right)^2$.
The uncertainty in $F$ due to errors in counting statistics was obtained by Monte Carlo simulation, where 36 new measurement results are numerically generated,
each drawn randomly from a Poissonian distribution with a mean equal to the original number of counts~\cite{altepeter2005photonic}. 
From these numerically generated results, a new density matrix can be calculated and consequently, a new value of $F$. 
Repeating this process 100 times, we obtain the fidelity distribution shown in Figure~\ref{fig:fidelity_distribution} from which we compute a 95\% confidence interval $98.7\%<F<98.9\%$. The distribution of $F$ does not include 100\%, 
which we attribute to imperfect control of the polarization state in the optical fiber. 
From the near-unity value of $F$, we conclude that the circulators do not affect the distributed Bell state.

\section{Conclusion}
We have  successfully demonstrated an attack of a clock synchronization protocol 
that tries to achieve security by detecting changes in 
polarization-entanglement distributed across a synchronization channel.
The attack was implemented by rerouting photons with polarization-insensitive circulators, and imposing a direction-dependent propagation delay. 
The observed shift in the estimated clock offset is equal to half the
propagation delay asymmetry, as expected for a protocol which assumes a symmetric channel~\cite{narula:17}.
Although circulators reroute photons using a polarization-rotation mechanism,  
we experimentally verify that they produce no measurable change in the distributed entangled state, 
indicating that they cannot be detected with the protocol. 

In this work, we focused on detecting its underlying mechanism -- Faraday
Rotation (FR), which must be performed in any circulator.
Methods based on characterizing light intensities,  e.g. identifying
additional reflections, may still allow the detection of circulators,
but they rely on the specific characteristics of the device (e.g. reflectivity). 
We also note that when Alice and Bob exchange photons that are identical in every other degree-of-freedom apart from propagation direction,
there are few technologies besides a FR-based circulator capable of discreetly separating their photons.
Alternatives such as advanced photonic
structures~\cite{jalas2014optical,dmitriev2013possible,dmitriev2013compact,bi2011chip,yu2009complete}
and quantum non-demolition measurements~\cite{lamas2018secure} still pose a
significant technological barrier for any adversary, so entanglement-based
clock synchronization still may provide a significant security advantage
compared to traditional methods.

In the supplementary material, we also examine the geometric phase associated
with polarization state rotation in the circulators, previously thought to be
observable~\cite{Troupe:2018}, as an additional phase associated with photon dynamics in the Faraday Rotator neutralizes this geometric phase. 
We note that when geometric phases were observed in other entangled systems, 
an interferometric arrangement was necessary to eliminate the influence of this ``dynamic'' phase~\cite{kwiat:91,strekalov:97,brendel:95,jha:08}. 
Whether or not a similar technique can be used to secure the present
synchronization protocol remains an open question. 

We acknowledge support by the National Research Foundation
\& Ministry of Education in Singapore. 
JT acknowledges support from the ARL:UT
Independent Research and Development Program.
\newpage
\bibliographystyle{apsrev4-1}
%

\newpage
\appendix
\section{Supplementary material}
In this section, we show that when circulators rotate 
the polarization state of one of the photons in an entangled pair by 180$^{\text{o}}$,
the geometric phase imposed on the rotated photon does not produce a measurable change in polarization entanglement.

We first introduce the formalism to deal with the fact that points on the Poincaré sphere carry no phase information;
the beginning and end points of a cyclic evolution correspond on the same point on the sphere.

To reflect this property, we define a ``basis vector field'' $\ket{\tilde{\psi}(t)}$, 
such that 
\begin{equation*}
\ket{\tilde{\psi}(t)}=e^{-if(t)}\ket{\psi(t)} \quad\text{and}\quad \ket{\tilde{\psi}(\tau)}=\ket{\tilde{\psi}(0)}, 
\end{equation*}
where $f(t)$ is the phase of $\ket{\psi(t)}$ expressed in terms of its basis state $\ket{\tilde{\psi}(t)}$ on the Poincaré sphere~\cite{anandan1992geometric}. 

The change in $f$ comprises of two terms
\begin{equation}\label{eqn:full_phase_change}
\Delta f = \beta + \gamma, 
\end{equation}
where the geometric phase
\begin{equation}
  \beta = \int\displaylimits_0^\tau\bra{\tilde{\psi}(t)}i\frac{\diff }{\diff t}\ket{\tilde{\psi}(t)}
\end{equation}
is due to the evolution of the basis state along a curved geometry, 
and the dynamic phase
\begin{equation}\label{eqn:dp_raw}
  \gamma = - \int\displaylimits_0^\tau \bra{\psi(t)}i\frac{\diff}{\diff t}\ket{\psi(t)} \diff t
\end{equation} 
is due to the photon's dynamics through the rotation medium~\cite{Troupe:2018}.
\subsection{Geometric Phase}\label{sec:gp}
Berry showed that the geometric phase is proportional only to the solid angle
$\Omega$ subtended by the cyclic trajectory on the Poincaré
sphere~\cite{berry:87},
\begin{equation}
\beta = -\frac{1}{2} \Omega.
\end{equation}
Thus, a qubit in the initial state
\begin{equation} \label{eq:qubit}
\ket{\psi(t=0)} = e^{-i\phi}\cos(\theta/2)\ket{R}+ \sin(\theta/2)\ket{L}
\end{equation}
that underwent a 180$^{\text{o}}$ rotation in the plane of polarization ($\phi\rightarrow\phi+2\pi$)
will accumulate a geometric phase $\beta = -\pi(1-\cos\theta)$.

\subsection{Dynamic Phase}\label{sec:dp}
To evaluate the dynamic phase $\gamma$ accumulated by the photon at end of a Faraday Rotator of length $d$, we parameterize its expression in Eq.~\ref{eqn:dp_raw} in terms of the penetration depth $z$
\begin{align}\label{eqn:dp}
\gamma &= -\int\displaylimits_0^d \bra{\psi(z)}i\frac{\diff }{\diff z}\ket{\psi(z)} dz
&=-\int\displaylimits_0^d \bra{\psi(z)}\hat{N}\ket{\psi(z)} dz,
\end{align}
where 
\begin{equation}\label{eqn:operator_and_state}
\hat{N}
    = k\begin{pmatrix} 
      n_R & 0 \\
      0 & n_L
    \end{pmatrix} \quad\text{and}\quad 
\ket{\psi(z)} = \begin{pmatrix} 
e^{ikn_Rz}e^{-i\phi}\cos(\theta/2) \\
e^{ikn_Lz}sin(\theta/2)
\end{pmatrix}
\end{equation}
are expressed in the $\{\ket{R},\ket{L}\}$ basis, and $k =
\frac{2\pi}{\lambda}$ is the wave number of the photon mode in free space.

The Faraday Rotator is a birefringent medium whose refractive indices $n_{R,L}$ depend on the magnitude of an applied magnetic field $B$ in the direction of light propagation,
\begin{equation}\label{eqn:refractive_index}
  n_{R,L} = n_0\left(1\pm\frac{VB}{kn_0}\right),
\end{equation}
where $V$ is the Verdet constant and $n_0$ is the index of refraction in the absence of a magnetic field.

Substituting \ref{eqn:operator_and_state} into \ref{eqn:dp}, we obtain
\begin{equation}\label{eqn:dp_psi}
  \gamma = kn_0d + VBd\cos\theta,
\end{equation}
where the product $VBd$
can be shown~\cite{zak1991geometric} to be the anti-clockwise rotation angle for a linearly polarized input.

Consider an initial input state 
$\ket{\psi(\phi=0,\theta=0)}=\ket{H}$. 
For the evolution cycle ($\phi = 0\rightarrow2\pi$) considered earlier,
$\ket{H}\rightarrow\ket{-45}\rightarrow{\ket{V}}\rightarrow\ket{+45}\rightarrow\ket{H}$ corresponds to a \textit{clockwise} 180$^{\text{o}}$ in the plane-of-polarization. 
Thus, the the rotation must be realized by a medium whose product $VBd = -\pi$.
Consequently, the dynamic phase $\gamma=kn_0d - \pi\cos\theta$ for the state considered in Eq.~\ref{eq:qubit}.

\subsection{Overall Phase \& the Circulator Attack}
We have already shown that an initial state 
\begin{equation}\label{eq:psi}
\ket{\psi} = \cos(\theta/2)\ket{R} + \sin(\theta/2)\ket{L}, 
\end{equation}
will accumulate a geometric phase $\beta = -\pi(1-\cos\theta)$ and a dynamic phase $\gamma = kn_0d - \pi\cos\theta$, resulting in an overall phase $\phi=kn_0d-\pi$.
Repeating this procedure for the orthogonal state
\begin{equation}\label{eq:psi_perp}
\ket{\psi_\perp} = -\sin(\theta/2)\ket{R} + \cos(\theta/2)\ket{L}, 
\end{equation}
we obtain a geometric phase of $\beta'=-\beta = +\pi(1-\cos\theta)$ and a dynamic phase $\gamma' = kn_0d + \pi\cos\theta$, resulting in an overall phase $\phi'=kn_0d+\pi= \phi+2\pi$.  

Let the entangled pair initially be in the Bell state $\ket{\Psi^-} = \frac{1}{\sqrt{2}}\left( \ket{HV} - \ket{VH}\right)$.  With the first qubit Alice's photon and the second one Bob's photon.  We can re-write the Bell state in the basis defined by  Equations \ref{eq:psi} and \ref{eq:psi_perp},
\begin{align}
\ket{\Psi^-} & =\frac{1}{\sqrt{2}} \Big( \ket{HV} - \ket{VH} \Big) \nonumber \\
 &= \frac{i}{\sqrt{2}} \Big( \ket{\psi_\perp}_A\ket{\psi}_B - \ket{\psi}_A\ket{\psi_\perp}_B \Big). 
\end{align}
The state of the Bell pair after Bob's photon goes through Eve's circulator based attack, $\hat{U}_{Attack}$, is given by 
\begin{align}
\ket{\Psi^-} 
& \rightarrow 
\hat{U}_{Attack}\frac{i}{\sqrt{2}}\Big( \ket{\psi_\perp}_A\ket{\psi}_B - 
\ket{\psi}_A\ket{\psi_\perp}_B \Big) \nonumber\\
&= 
\frac{i}{\sqrt{2}} \Big( 
e^{i\phi}\ket{\psi_\perp}_A\ket{\psi}_B 
- e^{i\phi'}\ket{\psi}_A\ket{\psi_\perp}_B \Big)\\
&=
\frac{ie^{i\phi}}{\sqrt{2}} \Big( 
\ket{\psi_\perp}_A\ket{\psi}_B 
- e^{i2\pi}\ket{\psi}_A\ket{\psi_\perp}_B \Big)\nonumber\\ 
&=e^{i\phi}\ket{\Psi^-}=-e^{ikn_0d}\ket{\Psi^-} \nonumber\\
&\equiv -\ket{\Psi^-}.
\end{align}
We can see from this expression, 
that the initial Bell state remains unchanged from the introduction of the circulators, 
and is equivalent to the result obtained by direct calculation in Eq.~\ref{eqn:direct_result}. 

Recent work assumed that the contribution from the dynamic phase was ``zero, or is known and compensated for'' and predicted instead that the circulators imparted a non-local geometric phase to produce a dramatic change~\cite{Troupe:2018}
\begin{align}
\ket{\Psi^-} 
& \rightarrow 
\hat{U}_{Attack}\frac{i}{\sqrt{2}}\Big( \ket{\psi_\perp}_A\ket{\psi}_B - 
\ket{\psi}_A\ket{\psi_\perp}_B \Big) \nonumber\\
&= 
\frac{i}{\sqrt{2}} \Big( 
e^{i\beta}\ket{\psi_\perp}_A\ket{\psi}_B 
- e^{-i\beta}\ket{\psi}_A\ket{\psi_\perp}_B \Big).
\end{align}
However, we note that the dynamic phase (Eq.~\ref{eqn:dp_psi}) is likewise non-local (due to its dependence on $\theta$) and combines with the geometric phase to produce no measurable net change in the state.
\end{document}